**Excess noise and electron-phonon energy transfer in Au wires: Length and temperature effects**

Yuxin Wan[1], Dale T. Lowder[1†], Liyang Chen[2‡], Manasi Mandal[1], Gavin Joyce[3], Douglas Natelson[1,3,4,5,6]

[1]*Department of Physics and Astronomy, Rice University, Houston, TX 77005 USA*
[2]*Applied Physics Program, Smalley-Curl Institute, Rice University, Houston, TX 77005 USA*
[3]*Department of Electrical and Computer Engineering, Rice University, Houston, TX 77005 USA*
[4]*Department of Materials Science and NanoEngineering, Rice University, Houston, TX 77005 USA*
[5]*Rice Center for Quantum Materials, Smalley-Curl Institute, Rice University, Houston, TX 77005 USA*
[6]*Rice Advanced Materials Institute, Rice University, Houston, TX 77005 USA*

**Abstract**:

As a mesoscopic diffusive wire at low temperatures is driven out of equilibrium by a current, excess noise is observed compared to the equilibrium Johnson-Nyquist fluctuations. The expected form of the excess noise varies depending on the relative rates of wire traversal, electron-electron inelastic energy transfer, and electron-phonon inelastic energy transfer. We measure noise as a function of bias current in diffusive Au nanowires on $SiO_2$/Si with lengths ranging from sub-μm to 30 μm and temperatures from 3 K to 50 K, to map out the interplay between charge shot noise and the crossover to electron self-heating and phonon-dominated noise suppression. As expected qualitatively, shot noise is suppressed with increasing wire length and substrate temperature, as electron-phonon scattering is of increasing importance. In long wires, the simple model of electron-phonon energy loss known to be quantitatively effective below a few Kelvin underpredicts the excess noise in the high bias, high temperature limit. This issue is more severe in wires on $SiO_2$/Si than on bare Si, implying that interfacial thermal conduction must be considered in the thermal path even at comparatively high temperatures.

## I. INTRODUCTION

Noise in electronic devices originates from fluctuations of current and voltage, with different microscopic processes and device geometries giving rise to distinct noise signatures. Noise measurements provide insights into electronic transport and inelastic scattering beyond conventional current-voltage ($I$ vs $V$) characterization. In thermal equilibrium, the electronic distribution function in a metal wire is the usual Fermi-Dirac distribution at a temperature $T$ that is identical to the lattice (phonon) temperature in the metal and the insulating substrate. This leads to Johnson-Nyquist noise[1,2] characterized by the voltage noise power spectral density (current noise power spectral density $S_V = 4k_BTR$ ($S_I = 4k_B T/R$), which remains "white" (frequency independent) as long as $k_BT \gg hf$, for a device of resistance $R$.

When a device is driven out of equilibrium by a current $I$, in the absence of extrinsic contributions such as flicker noise[3], the voltage (current) fluctuations generally increase. "Excess" noise is the voltage or current noise at finite bias above the equilibrium JN noise found at zero bias. As summarized in detail below, in diffusive wires short compared to the length scale associated with inelastic exchange of electronic energy with the phonons, the excess current noise, "shot noise"[4-6], increases at high biases ($eV \gg k_BT$), where $e$ is the electronic charge and $V$ is the voltage across the wire) linearly in the current $I$. In the high bias limit, $S_I \sim F \cdot 2eI$, where $F$ is the Fano factor. When electron-electron (e-e) inelastic scattering is weak, in a diffusive wire[7,8] $F \rightarrow 1/3$. When e-e scattering is rapid, so that a local electronic temperature profile $T_e(x)$ is established along the wire position $x$, then $F \rightarrow \sqrt{3}/4$.[9-11] When the electron-phonon inelastic scattering is strong, the additional current noise becomes suppressed. In the limit that the electrons are instantly thermally equilibrated with the phonons, $T_e(x) \approx T_{ph}$, and the current noise would be $S_I = 4k_BT_{ph}/R$. This is essentially the reason that shot noise is not routinely detectable in macroscopic conductors at room temperature. In the intermediate situation, $T_e(x)$ is elevated above $T_{ph}$ in the metal, so that noise as a function of bias in long wires (where $T_e(x)$ is expected to be almost uniform) provides a means of assessing the strength of the electron-phonon inelastic scattering.[11-13]

In this paper, we measure noise in a series of diffusive gold nanowires with varying dimensions over the temperature range 3 K- 50 K to map out the suppression of shot noise and the crossover to the electron-phonon dominated regime as a function of bias, temperature, device length. In the shortest nanowires with $\ell \ll L \ll L_{ee}$, we observe shot noise at 3 K with a Fano factor close to the expected universal value $F = 1/3$. The shot noise is suppressed below this value as wire length increases and temperature increases, as expected qualitatively, though detailed quantitative modeling is challenging as local electronic temperature is not necessarily well-defined. For long (30 μm) wires on $SiO_2$/Si, as $T$ is increased, the noise as a function of bias is larger and follows a different functional form than expectations for a fixed coupling to a phonon bath at the cryostat temperature. A candidate explanation is that the temperature of the phonons in the wire at high dissipated powers is elevated above those of the substrate. As a result, there is a greater bias-driven-heating contribution to noise at high temperatures than naively expected from extrapolating the low $T$ response. Comparing with a nominally identical Au nanowire fabricated on pure Si shows that the presence of the $SiO_2$ layer comparatively increases this bias-driven noise.

## II. BACKGROUND

In 1928, J.B. Johnson and H. Nyquist quantified and explained what is now called Johnson-Nyquist (JN) noise, generated by the thermal agitation of the charge carriers in a conductor at thermal equilibrium[1,2]. JN noise is characterized by the voltage noise power spectral density (current noise power spectral density) $S_V = 4k_BTR;\ S_I = 4k_B T/R$, which is "white" (frequency independent) as long as $k_BT \gg hf$, and $R$ is the real part of the device resistance. Johnson-Nyquist noise is a universal equilibrium property of conductors: as a direct consequence of the fluctuation-dissipation theorem, it is independent of the nature of the microscopic nature of charge carriers and depends only on the conductor's resistance and temperature.

"Excess" noise is the voltage or current noise at finite bias above the equilibrium JN noise found at zero bias. An intrinsic contribution is shot noise, which serves as a crucial experimental probe that effectively examines how charge flows through a material. Charge shot noise was first described by Walter Schottky in 1918 in studies of current fluctuations in vacuum tubes[4]. It is the time-dependent fluctuation of the electrical current in electronic circuits that arises from the discrete nature of charge carriers and the statistical variations in their arrival times following transport through the device. Measurements of shot noise have since revealed the underlying charge dynamics in a wide range of transport phenomena[6]. As a nonequilibrium phenomenon, shot noise can appear in a conductor when the electronic distribution is driven away from its thermal equilibrium form.

There are multiple time scales and hence length scales relevant to this problem. In a good metal at low temperatures, the shortest of these is the elastic mean free path for charge carriers, $\ell$, set by static disorder, including grain boundaries, interfaces, vacancies, and impurities. For wires shorter than $\ell$, electronic transport is ballistic, and shot noise is suppressed[5,14].

When $\ell$ is small compared to the wire dimensions (length $L$, width $w$, thickness $d$), the wire is in the diffusive regime. In the limit of no electron-electron inelastic scattering, the distribution function in the middle of such a wire has a two-step structure, as the applied bias shifts the chemical potential of carriers from the source contact relative to those from the drain contact by $eV$. Calculations of the current-current correlation function based on this distribution function and the Boltzmann formalism find shot noise with $F = 1/3$ in this limit[8]. Equivalently[7], treating a diffusive conductor as the limit of elastic scattering of specific quantum channels of different transmittances, the same result is found in the absence of e-e scattering.

There is also a length scale $L_{ee}$ associated with the inelastic e-e scattering that redistributes energy among the electrons and smooths out what would otherwise be a non-thermal electronic distribution. This energy relaxation length scale has been examined in detail below 1 K by experiments using tunneling measurements to infer the local distribution function in a quasi-1D wire as a function of temperature and bias conditions[15,16]. A calculation based on e-e inelastic scattering as a source of decoherence for quantum corrections to conduction gives[17], for a quasi-1D wire,

$$L_{ee} = \left[\left(\frac{\sqrt{2}}{k_B}\right)\left(\frac{\hbar}{e}\right)^2 \frac{Dw}{TR_\square}\right]^{1/3} \qquad (1)$$

Here $D$ is the electronic diffusion constant, $R_\square$ is the resistance per square of the wire, and the quasi-1D requirement is that $L_{ee}$ exceeds $w$ and $d$. Empirically, this expression overestimates the length scale for redistributing the energy throughout the electronic distribution[18,19]. Scattering from magnetic impurities can also shorten that length scale, particularly below 1 K[20], and Eq. (1) still appears to be an overestimate in highly pure materials[18,19].

If we assume that the e-e scattering process is much faster than inelastic electron-phonon scattering, one can define a locally thermal Fermi-Dirac distribution characterized by a locally elevated electronic temperature $T_e(x)$. Assuming that the ends of the wire are thermally anchored at the substrate temperature and that electronic thermal conduction consistent with Wiedemann-Franz takes place out of the well-heat-sunk ends of the wire, the result for $T_e(x)$ is a parabolic profile[10]. The current noise then becomes $S_I = \frac{4k_B\langle T_e(x)\rangle}{R}$, where $\langle T_e(x)\rangle$ is the length-averaged local electron temperature, assuming a temperature-independent resistivity $\rho \approx \rho_0$ for these low temperature measurements. Because $\langle T_e(x)\rangle \propto I$ in this limit, the excess noise in the strong e-e scattering with local thermalization looks like shot noise with a Fano factor of $\sqrt{3}/4$. Observing this Fano factor is quite challenging experimentally[11], requiring large, thick contact pads to ensure that the wire ends are properly thermally anchored.

Electron-phonon scattering is important for longer wires and elevated temperatures. In the limiting case that the electrons are very well thermally anchored to the phonons, and the phonons in the wire are equilibrated at the cryostat temperature, $T_0$, then $T_e \approx T_{ph} = T_0$, and there is no excess noise. Assuming instead that the electrons at each location $x$ along the wire have a Fermi-Dirac distribution characterized by a local $T_e(x)$, a simple model of electron-phonon inelastic scattering leads to the prediction[12,21] that the rate of energy loss from the electrons to the phonons scales like $T_e^5(x) - T_{ph}^5$ in the low temperature limit. The basis for this is an e-ph scattering rate[22] that scales like $T_e^3$, which assumes (i) 3D phonons; (ii) $T_e, T_{ph} \ll T_D$, the Debye temperature; and (iii) $T_e \ll T_\mathrm{F}$, the Fermi temperature. This has proven a good description of a number of experiments below 4 K[19]. In the "dirty" limit $\lambda_{ph} \gg \ell$, (the thermal phonon wavelength $\lambda_{ph}(T_{ph}) \sim hc_s/k_B T_{ph}$, where $c_s$ is the speed of sound in the metal, typically several km/s, and $\ell$ is the elastic mean free path for electrons), other temperature dependences are possible, depending on the nature of the disorder[23,24]. The energy transfer rate can vary like $T_e^4$ or $T_e^6$ (for quantum interference in the presence of static disorder, or vibrational dragging of the disorder potential, respectively).

Within the standard $T_e^5 - T_{ph}^5$ approach, the temperature profile can be found by solving

$$\frac{\pi^2}{6}\frac{d^2T_e^2}{dx^2} - \frac{\pi^2 T_e}{3r}\frac{dr}{dT_e}\left(\frac{dT_e}{dx}\right)^2 = -\left(\frac{eE}{k_B}\right)^2 + \Gamma(T_e^5 - T_{ph}^5) \qquad (2)$$

where $r(T_e)$ is the resistance per unit length, $E$ is the local electric field within the wire, and $\Gamma$ parametrizes the energy transfer rate between electrons and phonons. The energy added to the electrons via Joule heating flows into the phonons and is also conducted along the length of the wire (assuming Wiedemann-Franz governs the electronic thermal conductivity). For the usual situation of low temperatures such that the resistance is approximately temperature-independent, this reduces to

$$\frac{\pi^2}{6}\frac{d^2T_e^2}{dx^2} = -\left(\frac{eE}{k_B}\right)^2 + \Gamma(T_e^5 - T_{ph}^5) \qquad (3)$$

In the absence of phonons ($\Gamma = 0$), this leads to the $F = \sqrt{3}/4$ current noise result mentioned above.

In this model, one can define an electron-phonon scattering length, $L_{e-ph} \approx 1.31/\sqrt{(T^3\,\Gamma)}$. In the limit that $L \gg L_{e-ph}$, the electronic temperature $T_e(x)$ is essentially uniform along the wire and elevated above the phonon temperature, $\langle T_e \rangle \approx \left(\frac{1}{\Gamma}\left(\frac{eV}{L}\right)^2 + T_{ph}^5\right)^{1/5}$, where the wire is assumed to be Ohmic. The voltage noise (current noise) is then the JN noise for this elevated temperature, $S_V = 4k_B\langle T_e\rangle R$ ($S_I = 4k_B\langle T_e\rangle/R$), which is excess noise as $\langle T_e\rangle$ ramps up with increasing bias. A comprehensive investigation of this model and temperature profiles across the full range of the competition between longitudinal electronic thermal conduction and energy transfer to phonons has been performed by Huard *et al.*[25], including looking at noise and the next highest cumulant of the current.

From fitting the measured noise in such a long wire as a function of bias, it is then possible to extract a value for $\Gamma$ within this model. This approach has worked well below a few Kelvin in a number of materials[19]. For $L = 50$ μm gold wires on oxidized Si measured at 2 K, Henny *et al.* reported $\Gamma \approx 5 \times 10^9$ $K^{-3}m^{-2}$ [13], and for $L = 30$ μm nanowires of the heavy fermion strange metal $YbRh_2Si_2$ at 3 K, $\Gamma \approx 9 \times 10^9$ $K^{-3}m^{-2}$ described the data well[26].

We note that at high substrate temperatures, when the magnitude of the inferred electron temperature elevation is small compared to $T_0$, it is difficult to discriminate between the $T^5$ model of electron cooling and possible alternative power laws. While the bulk $T_D$ for Au is 165 K[27], suppression of the Debye temperature to between 83 and 121 K has been inferred in Au thin films[28]. Measurement temperatures in this work are still well below this, but some change in e-ph cooling mechanism is conceivable at sufficiently high temperatures.

The crossover situation in shorter wires, when the electronic distribution function is not described by a locally thermalized form yet is strongly perturbed by e-e and e-ph inelastic processes, is generally complicated[29]. Given a model for the position- and energy-dependent distribution function, $f(\epsilon, x)$ (which depends implicitly on $T_{ph}$), current noise can be computed from the current-current correlation function via the approach of Nagaev[8,9], which leads to

$$S_I = \frac{4}{RL}\int_0^L dx \int d\epsilon\, f(\epsilon, x)[1 - f(\epsilon, x)] \qquad (4)$$

Arriving at a form for $f(\epsilon, x)$ in general requires solving a Boltzmann-like equation with scattering integrals that account for both e-e and e-ph processes[9]. Detailed considerations make a general treatment difficult, and uniqueness is a challenge due to multiple adjustable parameters.

As the cryostat temperature $T_0$ increases, a naive expectation is that the model of Eq. (2, 3) should become a better description of the noise in a nanowire. The characteristic thermal phonon wavelength $\lambda_{ph}(T)$ becomes shorter like $1/T$, so that the "dirty limit" $\lambda_{ph}(T) \gg \ell$ is readily avoided. Thermal boundary resistances for phonons tend to decrease as $1/T^p$, where $p$ is typically 3 for 3D phonons[30,31] (though this dependence is not as steep above a few Kelvin[30]). As $T_0$ increases, the phonons ($T_{ph}$) in the wire should naively be better thermally anchored to the substrate, and the substrate should be better thermally anchored to the cryostat.

## III. FABRICATION AND MEASUREMENT

Nanowires are fabricated through multi-step electron beam lithography and liftoff processing with lengths ranging from submicron to several micrometers, each connected to two thick Au contact pads on substrates comprising 300 nm thermally grown $SiO_2$ on Si. A double-layer resist process using polymethyl methacrylate (PMMA) improves yield and edge quality. The nanowires are patterned to be 100 nm in nominal width using an Elionix G-100 e-beam lithography system, followed by development and an Ar plasma descum. A 1 nm Ti adhesion layer followed by an 18 nm Au layer is then deposited by electron beam evaporation. The lift-off process is performed using warm acetone, leaving the Au nanowires on the oxide surface. Two 300 μm × 300 μm electrical contact pads are patterned onto the Au nanowires, and a 2 nm Ti / 200 nm Au layer is deposited via e-beam evaporation and liftoff as above. Patterning the wire and the contact pads separately avoids issues with e-beam proximity effects. Scanning electron microscopy images below confirm well-defined nanowires with widths of approximately 100 nm. Large area, thick contact pads are necessary to thermally anchor the ends of the wires and avoid erroneous extra contributions to the noise from heating of the contacts[11].

Noise measurements are performed using a home-built setup capable of both biasing the device and collecting the voltage noise signal (Figure 1). The basic broadband cross-correlation approach is well established[11,32,33] and reduces the vulnerability of the measurement to the amplifier noise. The device is mounted on a custom-built low-frequency measurement probe electrically isolated from the cryostat (Quantum Design Physical Property Measurement System) ground. A heavily filtered National Instruments DAQ is used to supply a dc voltage across symmetrically arranged large resistors (each 220 kΩ), to act as a source of essentially constant average dc current. Voltage noise across the device arising from the current fluctuations is captured using two parallel amplifier chains (each consisting of a NF Electronics LI-75A voltage preamplifier (100× gain) and a Stanford Research SR560 voltage preamplifier (100× gain)). At each bias condition, the time-dependent voltage is sampled using a high-speed Picoscope 4262, with each sample consisting of 2 million data points at a sampling rate of 10 MHz. The Fourier transformed cross-correlation between output voltages of the two amplifier chains is computed for 300 samples and the results are averaged. This is repeated three times at each bias condition and the results are averaged. Breaking up the measurements allows discrimination against occasional glitches that could otherwise disrupt long, continuous averaging. For higher resistance wires there can be a drop off in detected noise at high frequencies due to $RC$ filtering effects from the capacitance of the wiring leading from the sample to the first-stage amplification. This is irrelevant in the present case, however, due to the low wire resistances and the bandwidth selected for the measurement.

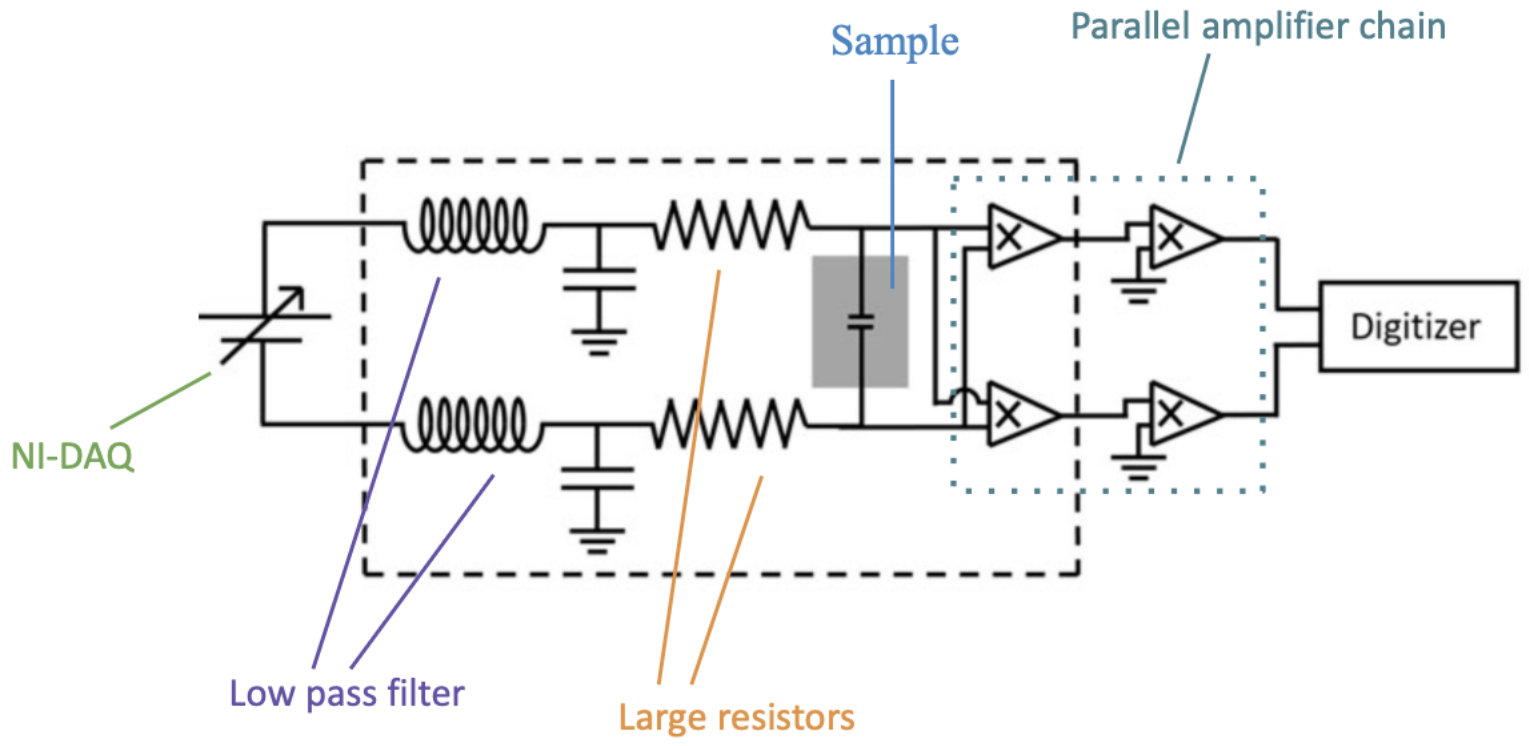

FIG 1. Schematic diagram of noise measurement configuration.

To determine the overall amplifier calibration, a 100 Ω metal film resistor was measured as a function of temperature to obtain a careful $R(T)$ vs. $T$. At zero applied bias, the Johnson-Nyquist noise was measured as a function of temperature. Different frequency bands are examined to avoid pickup of extrinsic noise sources. The frequency band with minimal extrinsic pickup is between 80 kHz and 300 kHz; narrower regions within this band are sometimes used to further avoid extrinsic noise peaks. The voltage noise power spectral density obtained through the cross-correlation is described by

$$S_{JN} = A \cdot 4k_B TR + S_{bg}, \qquad (5)$$

where $A$ is the effective gain of the amplification chain and cross-correlation procedure and $S_{bg}$ is the residual background noise of the setup, which ideally would be zero if the cross-correlation procedure removed all the amplifier noise and introduced no additional contributions. As shown in Figure 2, the measured noise scales linearly with $T \cdot R$. From a linear fit to these data, we extract $A = 7.768 \times 10^8$ and $S_{bg} = 1.012 \times 10^{-11}$ V$^2$/Hz; these values are used in the subsequent analysis of the nanowire noise. The inferred voltage noise power spectral density in the device, $S_V$, is related to the measured voltage noise power spectral density from the cross-correlation, $S_{V,meas}$, via $S_V = (S_{V,meas} - S_{bg})/A$. This resistor and all the wires were measured in a quasi-4-terminal configuration (two leads on source, two leads on drain), such that differential resistance of the cryostat wiring was irrelevant; however, differential resistance and noise measurements are all two-terminal, in the sense that pad and contact contributions add to those from the wires themselves. Contact resistances in similarly made structures are low (~ Ohm-level).

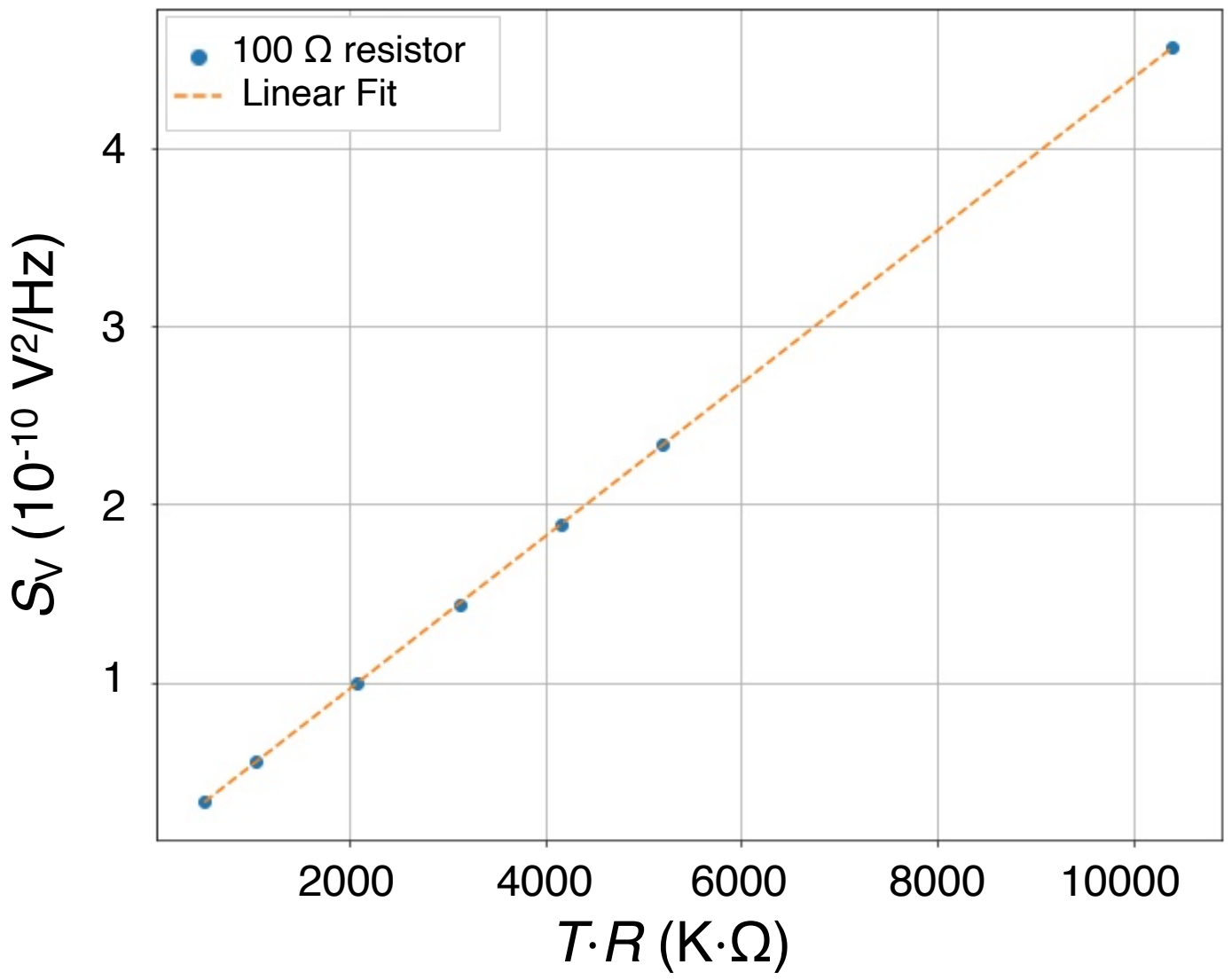

FIG. 2. Amplifier calibration data set, using measured voltage noise (including gain) for a 100 Ω metal film resistor as a function of cryostat temperature and the measured resistance at each cryostat temperature.

Flicker noise due to temporal fluctuations in the device resistance, $\delta R(t)$, can be a concern at non-zero bias current $I$. The voltage noise power for flicker noise scales like $\left(I\ \delta R(t)\right)^2 \sim I^2$, and often has a frequency dependence of the form $1/f^{\alpha}$, where $\alpha$ is close to one[3,34]. While some amount of flicker noise can be seen in Au nanostructures below a few Hz[35,36], this is negligible in the present experiments at the bandwidths employed here.

## IV. ANALYSIS, RESULTS, AND DISCUSSION

In the experiment, the cross-correlation setup measures voltage noise power spectral density, $S_V$. To focus on the excess noise, at a given cryostat temperature we plot the difference between the noise at finite bias current and the noise at zero bias. For all the wires measured, we measure the differential resistance $dV/dI$ as a function of the bias current via standard lock-in techniques. As shown in the example in Fig. 3, there are small corrections from the zero-bias resistance due to current-driven heating. The differential resistance is relevant in the analysis of the data when converting between the measured voltage noise $S_V$ and the current noise $S_I$, and for determining the voltage drop, $V$, across the wire.

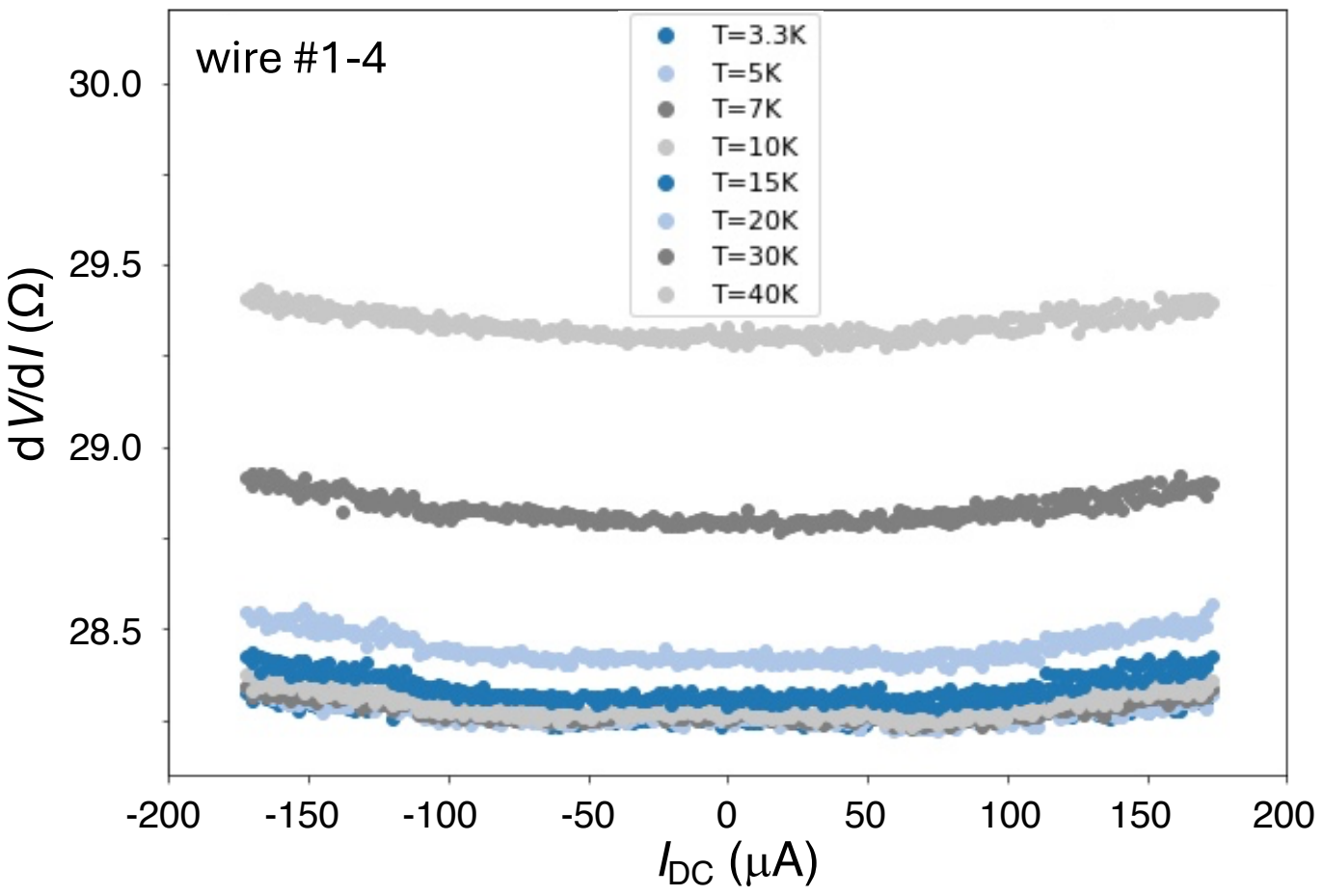

FIG. 3. Example data for the differential resistance as a function of bias current for wire 1-4 (448.5 nm in length), acquired via lock-in measurements at 7.7 Hz. The small apparent deviations from Ohmic response result from heating. Results for the other wires are similar.

The measured excess voltage noise can be analyzed in terms of the different models for its production. In the context of shot noise, intrinsically due to current fluctuations, the noise is modeled in terms of a Fano factor:

$$S_I = F \cdot 2eI \coth\left(\frac{eV}{2k_BT}\right) + (1-F)4k_BT\left(\frac{dV}{dI}\right)_I^{-1} \qquad (6)$$

The excess voltage noise is then

$$S_{V,excess}(I) = \left(\frac{dV}{dI}\right)_I^2 \left[F \cdot 2eI \coth\left(\frac{eV}{2k_BT}\right) + (1-F)4k_BT\left(\frac{dV}{dI}\right)_I^{-1}\right] - 4k_BT\left(\frac{dV}{dI}\right)_{I=0} \qquad (7)$$

Here $T$ is the zero-bias electronic temperature, which is expected to be $T_0$ . This expression holds even in the hot-electron, no phonon scattering case when $F$ is expected to be $\sqrt{3}/4$ and the local electron temperature is elevated. One can always analyze the excess noise in this framework, assigning an effective $F$ to any noise vs. bias current data set, for comparison to shot noise expectations.

The other limiting case is in the context of electron-phonon coupling dominating the noise, as in the model of Eq. (3). The voltage noise power spectral density as a function of bias at a given $T_0$ is examined by varying the parameter $\Gamma$, solving Eq. (3) to compute a temperature profile $T_e(x)$, and then finding the expected JN noise, $S_V = 4k_B\langle T_e\rangle(I)\left(\frac{dV}{dI}\right)_I$, where $\langle T_e\rangle(I)$ is the bias-dependent average electronic temperature. The excess noise is then

$$S_{V,excess}(I) = 4k_B\left[\langle T_e\rangle(I)\left(\frac{dV}{dI}\right)_I - T_0\left(\frac{dV}{dI}\right)_{I=0}\right]. \quad (8)$$

A best-fit algorithm can be run to determine what value of $\Gamma$ best aligns this model with the data. Again, this calculation can always be performed, and systematic disagreement of the data with the functional form obtained from this calculation would imply that the model of Eq. (3) is not adequate for that data set.

We fabricated four comparatively short Au nanowires with different dimensions in the regime where at low temperatures shot noise is expected to be a dominant contributor to the excess noise (Figure 4), as well as a ~ 30 μm long wire. The wires were deposited in two batches (wires 1-4 and 1-12 codeposited; wires 3-8, 3-11, and the long wire codeposited). The SEM images show the short wire dimensions. The long wire gives a good measure of the average resistivity of the gold in that deposition, with a zero-bias resistance at 3 K of 635 Ω, a length of 29.8 μm, and a width in SEM of 100 nm. From this, we can infer for the 18 nm thick Au films $\rho = 3.83$ μΩ-cm. Knowing that the density of states at the Fermi level for Au is approximately $\nu(\epsilon_F) = 1.0 \times 10^{47}$ $J^{-1}m^{-3}$, we can use the Einstein relation, $\sigma = e^2 D\nu(\epsilon_F)$ to estimate the electronic diffusion constant, $D \approx 1.01 \times 10^{-2}$ $m^2/s$. This implies an elastic mean free path $\ell$ ~ 14-21 nm, at the crossover between 2D and 3D diffusion ($\ell$ is comparable to film thickness). From Eq. (1), this implies an e-e scattering length $L_{ee} \approx 1.92$ μm at 3 K. For comparison, the 30 μm long Au nanowire from Ref. [26] fabricated on bare (native oxide only) Si, of similar dimensions, has a zero-bias resistance at 3 K of 525 Ω. Wire dimensions and two-terminal zero-bias resistances at 3 K are compiled in table I. We note that the comparatively higher two-terminal resistances for wires 1-4 and 1-12 are very likely due to higher resistivity for that deposition, as explained further below.

Table 1. Wire dimensions and two-terminal resistances. Dimensions inferred from SEM images in Fig. 4. Wires 1-4 and 1-12 were co-evaporated; wires 3-8 and 3-11 were co-evaporated in a separate deposition.

| Wire identifier | Length (± 2 nm) | Width (± 2 nm) | R @ 3 K |
|---|---|---|---|
| 1-4 | 449 nm | 101 nm | 28.24 Ω |
| 1-12 | 1.623 μm | 101 nm | 54.99 Ω |
| 3-8 | 768 nm | 79 nm | 30.46 Ω |
| 3-11 | 2.752 μm | 86 nm | 69.43 Ω |
| 30 μm on $SiO_2$/Si | 29.8 μm | 100 nm | 635 Ω |
| 30 μm on bare Si [26] | 30.0 μm | 100 nm | 525 Ω |

The voltage noise power spectral density was found using the bandwidth between 80 kHz and 300 kHz, a range comparatively free of pickup from extrinsic interference. Over this bandwidth, the measured voltage noise is frequency independent, as expected. Figure 5 shows $S_{V,excess}$ as a function of bias for several temperatures for wire 1-4, plotted with both the theoretical curve for the pure shot noise $F = 1/3$ result at finite temperature, and the curve for the best-fit Fano factor. As $T_0$ is increased, the curve for the $F = 1/3$ decreases in magnitude over the accessible bias current range because of the behavior of coth $\left(\frac{eV}{2k_BT}\right)$. The measured excess noise decreases more rapidly than this, as shot noise is suppressed due to electron-phonon scattering. This analysis was performed assuming negligible contact resistance. If instead the measured differential resistance includes a contact resistance contribution $R_c$, and the voltage drop across the actual wire is smaller than assumed (by $IR_c$), then the $F$ extracted from our present analysis would systematically underestimate the true Fano factor. (For wires 1-4, if there was a contact resistance of several Ohms, the inferred true Fano factor would be unphysically large, exceeding 1. This is further evidence that the contact resistances must be small compared to the total two-terminal resistance.)

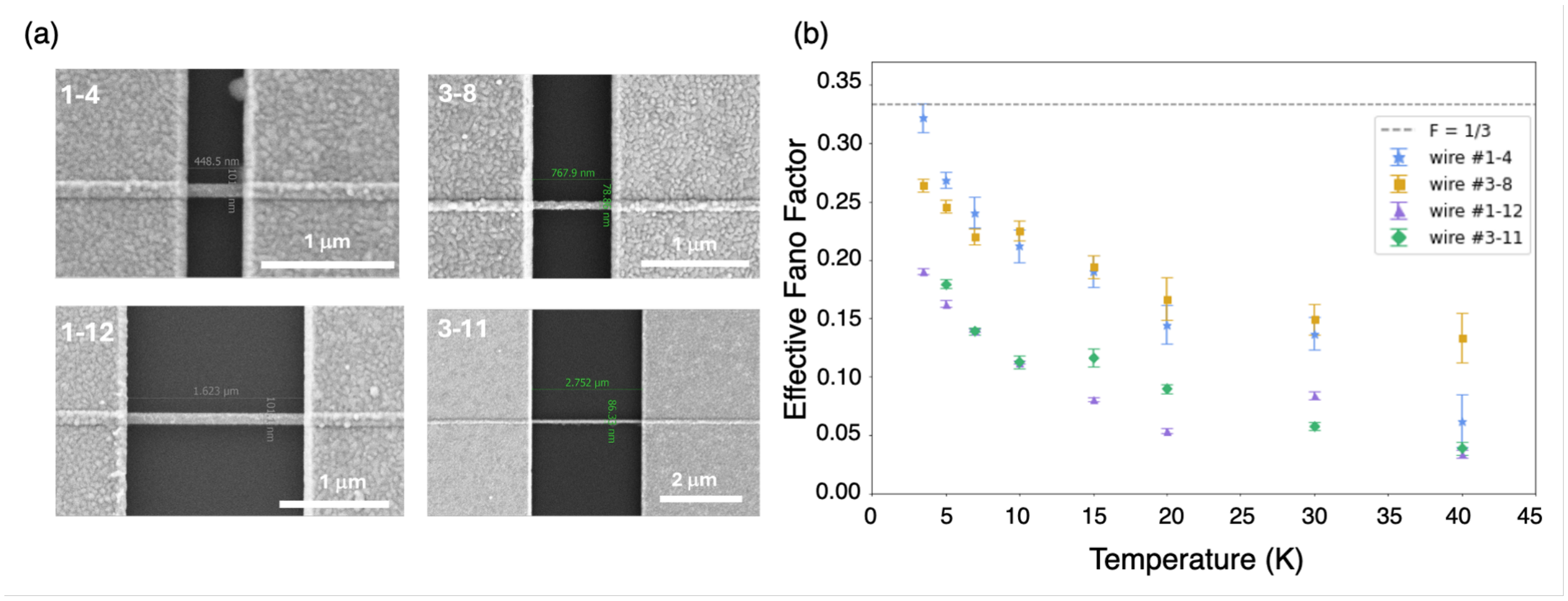

FIG 4. Various length Au nanowires and excess noise suppression as a function of temperature. (a) Scanning electron micrographs of four Au nanowires of various geometric dimensions, all contacted by large Au pads on $SiO_2$/Si substrates. The device dimensions are given in Table 1. (b) Effective Fano factor (from fitting to Eq. (7)) as a function of cryostat temperature for the four devices, illustrating the progressive suppression of excess noise with increasing temperature and wire length. Error bars (one sigma) are found from $\chi^2$ with one degree of freedom[37] for the fitting to Eq. (7) at each bias, with Fano factor as the sole fitting parameter.

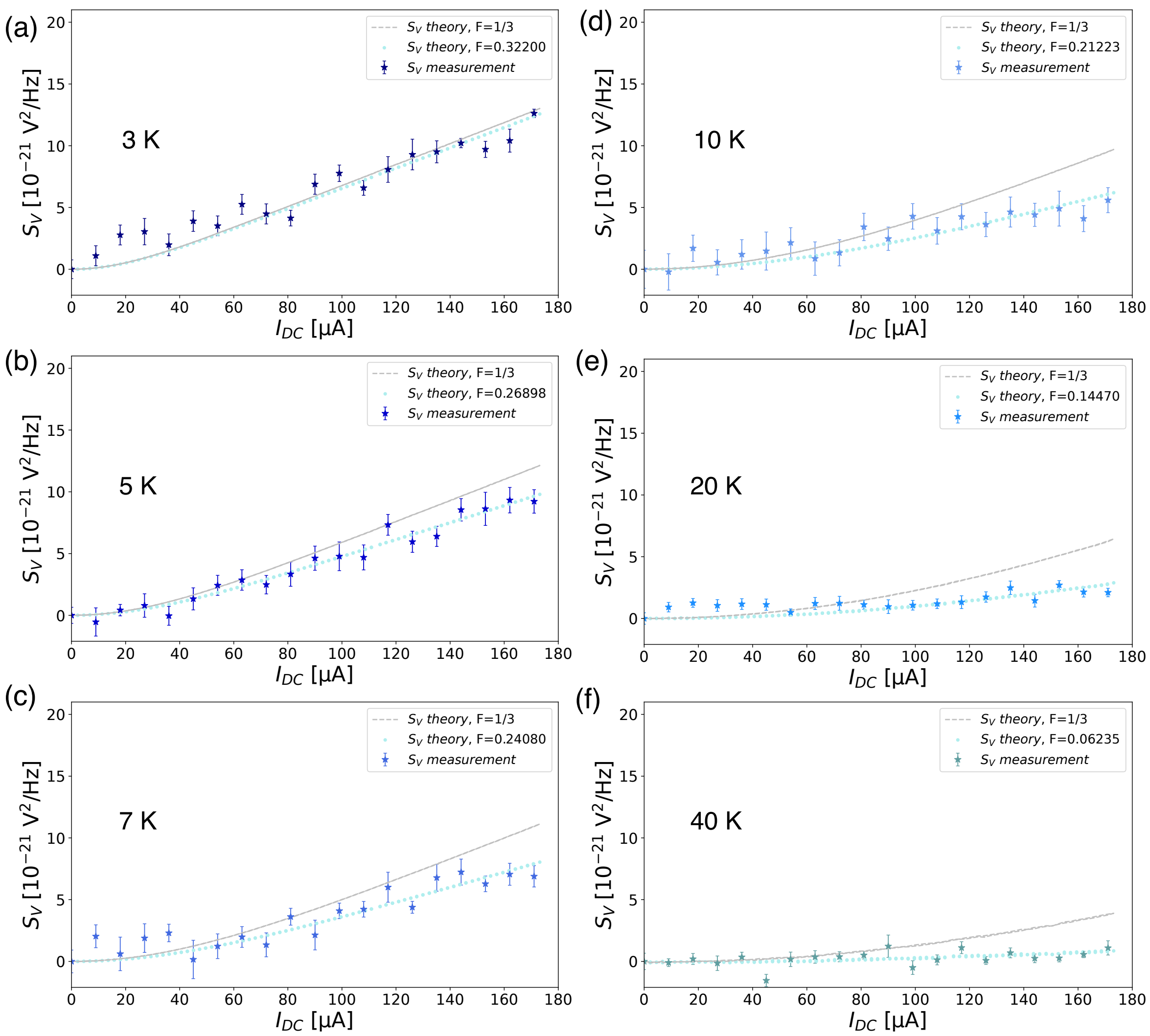

FIG 5. Wire #1-4, the voltage noise spectral density $S_V(I_{DC})$ for bath temperatures from 3 K to 40 K. Symbols show the measured data, and the sky-blue curves fit the shot-noise model of Eq. (7). As the temperature increases, the excess noise is suppressed as the $S_V$ $(I_{DC})$ curves flatten faster than what would be expected for pure shot noise at a fixed Fano factor of 1/3. The color scale runs from navy (3 K) to light blue (40 K). Error bars on each data point are from the standard error of the averaging procedure in finding the mean voltage noise.

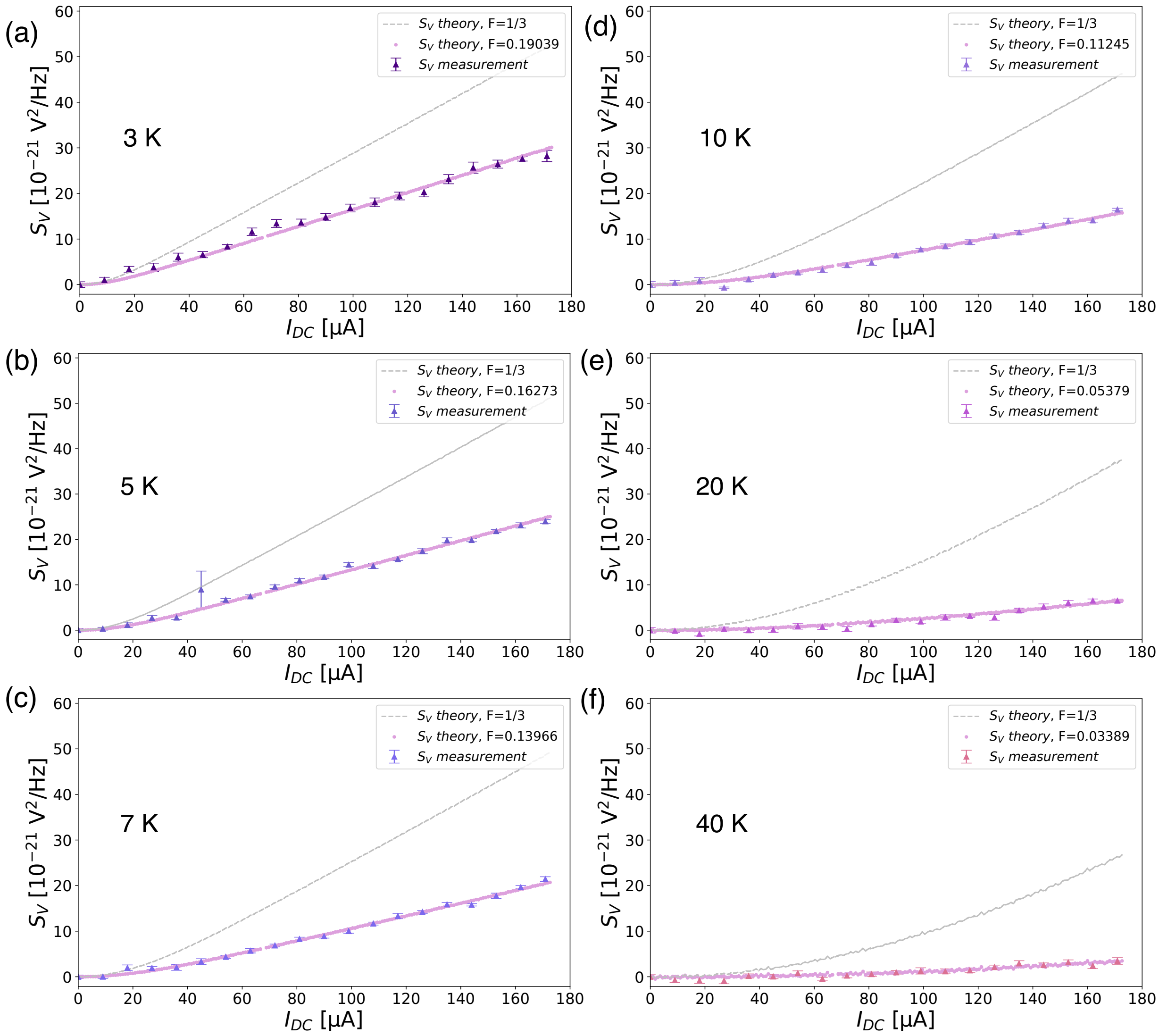

FIG 6. Wire #1-12, the voltage noise spectral density $S_V(I_{DC})$ for bath temperatures from 3 K to 40 K. The color scale runs from purple (3 K) to pink(40 K). Data points with error bars are measured values. The gray dashed line is the expected result for pure shot noise with $F = 1/3$.

Figure 6 shows the equivalent data sets for wire #1-12, $L \approx 1.62$ μm. As is evident, even though the device length is about half of the expected $L_{e-ph}$, the noise at 3 K is already suppressed below the $F = 1/3$ prediction. The suppression is stronger as the temperature is increased.

These data are taken throughout a crossover regime when the wires are not vastly longer compared to $L_{ee}(T_0)$. With the parameters mentioned above, $L_{ee}$ is estimated from Eq. (1) to be around 0.8 μm at the highest temperature of 40 K, while $L_{e-ph} \approx 73$ nm for that temperature, assuming for gold $\Gamma = 5 \times 10^9$ $K^{-3}m^{-2}$. Note that such an estimate for $L_{e-ph}$ is not obviously trustworthy, given that it is based on Eq. (3), which assumes that there is a well-defined local $T_e(x)$. Empirically, the data in Fig. 5 and 6 are well-fit by the expression in Eq. (7), which assumes shot noise with Fano factor as the adjustable parameter. This suggests that in this crossover regime at low cryostat temperatures the local distribution function in comparatively short wires remains locally non-thermalized in the measured substrate temperature range, but that shot noise is indeed reduced by coupling to the phonons. At elevated temperatures (20 K and above), the excess noise is sufficiently small that such qualitative agreement with the functional form of Eq. (7) may be fortuitous.

We also examine the noise in much longer (30 μm) Au wires, which should unambiguously be in the limit that $L \gg L_e, L_{e-ph}$ across the whole temperature range examined. We measure two such wires, one fabricated on 300 nm $SiO_2$/Si and one fabricated on "bare" (native oxide only) Si. Fig. 7 shows the measured excess voltage noise as a function of bias at both 3 K and 30 K for the $SiO_2$/Si substrate, and 3 K and 40 K for the pure Si substrate, respectively.

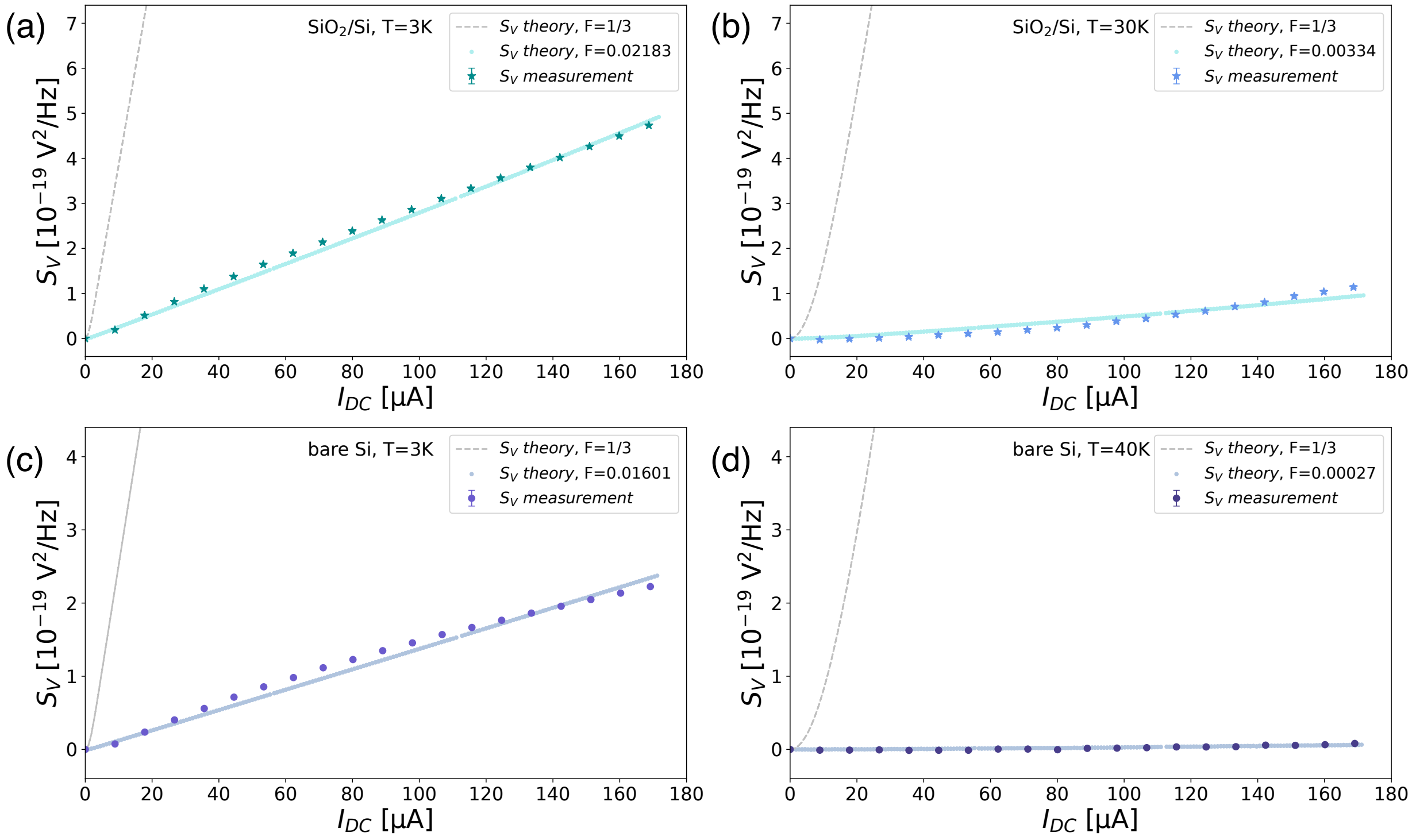

FIG 7. Excess noise vs. bias for long Au wires on different substrates. (a, b) $S_{V,excess}$ as a function of bias current for a 30 μm long Au wire on $SiO_2$/Si. (c, d) $S_{V,excess}$ as a function of bias current for a 30 μm long Au wire on nominally bare Si (just native oxide). Purely for illustrative purposes, the gray dashed lines show the expected noise if the samples were in the shot noise regime with $F = 1/3$, which is obviously not the case. Similarly, the solid lines are fits to Eq. (7), purely to get an illustrative sense of how reduced the noise is in this electron-phonon dominated regime relative to the shot noise case. The data points are the measured (Johnson-Nyquist) excess noise from the bias-heated electrons. Error bars (comparable to symbol sizes) are from standard error of the averaged data. Subsequent figures show the data for comparison to e-ph heating models.

In these devices, as expected, the excess noise is greatly suppressed relative to the magnitude that would be expected if one naively calculated the shot noise. Misapplying Eq. (7) to fit the measured data, the inferred Fano factors are ~ 0.02 at 3 K for both devices and decrease further with increasing $T_0$. The noise at 3 K in both devices has systematic deviations from the functional form of Eq. (7), as expected. Note that the noise in the $SiO_2$ substrate device is noticeably systematically larger than that in the bare Si device.

Having established that the long nanowire operates in the strongly thermalized regime with $F \rightarrow 0$, we analyze the excess noise in terms of the electron-phonon thermalization model of Eq. (3) and Eq. (8). In the idealized version of this model, a single value of $\Gamma$ should lead to an optimized fit of noise vs. bias at all temperatures. Figure 8 shows, for the $SiO_2$/Si substrate at each temperature, the measured excess noise as a function of bias (along with the knowingly inapplicable shot noise fit) as well as calculated noise vs. bias expectations from Eqs. (3),(8) for different values of $\Gamma$, and accompanying calculated temperature profiles (from Eq. (3)). Analogously, Fig. 9 contains the equivalent plots for the bare Si substrate case.

For the $SiO_2$/Si substrate case, the functional form for $S_{V,excess}(I)$ from the electron thermalization with phonons model does not agree well with the observed excess noise's bias dependence for *any* of the temperatures. At 3 K, for example, the "best fit" $\Gamma \approx 2 \times 10^9$ K$^{-3}$m$^{-2}$ only roughly threads the data, overshooting the measurements at lower biases and undershooting at higher biases. The model predicts a pronounced negative curvature with bias at high biases, while the data themselves are much better approximated by a roughly linear-in-$I$ trend at high biases. This is consistent with a phenomenological form $\langle T_e \rangle \approx \left(T_0^2 + a(IV)\right)^{1/2}$ , for an Ohmic conductor. Such a dependence was discussed in Ref. [11] in the context of the entire substrate being driven to elevated temperatures due to poor thermal contact between the substrate and the cryostat. That is a serious concern below 4.2 K because of thermal boundary resistances and their strong temperature dependence. However, at higher temperatures, thermal contact between the substrate and the cryostat should improve significantly. As $T_0$ increases, the fits to the functional form of the thermalized model (which assumes $T_{ph} = T_0$ in the model's simplest form) improve qualitatively, but quantitative issues remain. At 20 K, the inferred $\Gamma \approx 3 \times 10^8$ K$^{-3}$m$^{-2}$ needed to match the data is an order of magnitude lower than the literature value and the 3 K parameter for this device. At 30 K, the inferred $\Gamma \approx 1.2 \times 10^8$ K$^{-3}$m$^{-2}$. At these elevated $T_0$ levels, the inferred $\Gamma$ should be considered an illustrative way to compare the excess noise, rather than a rigorous statement about the electron-phonon energy loss. Rather, within the low-$T$ model of electron thermalization with phonons, the excess noise is consistently larger than what would be expected for the established $\Gamma$ for gold, and this issue persists up to tens of Kelvin.

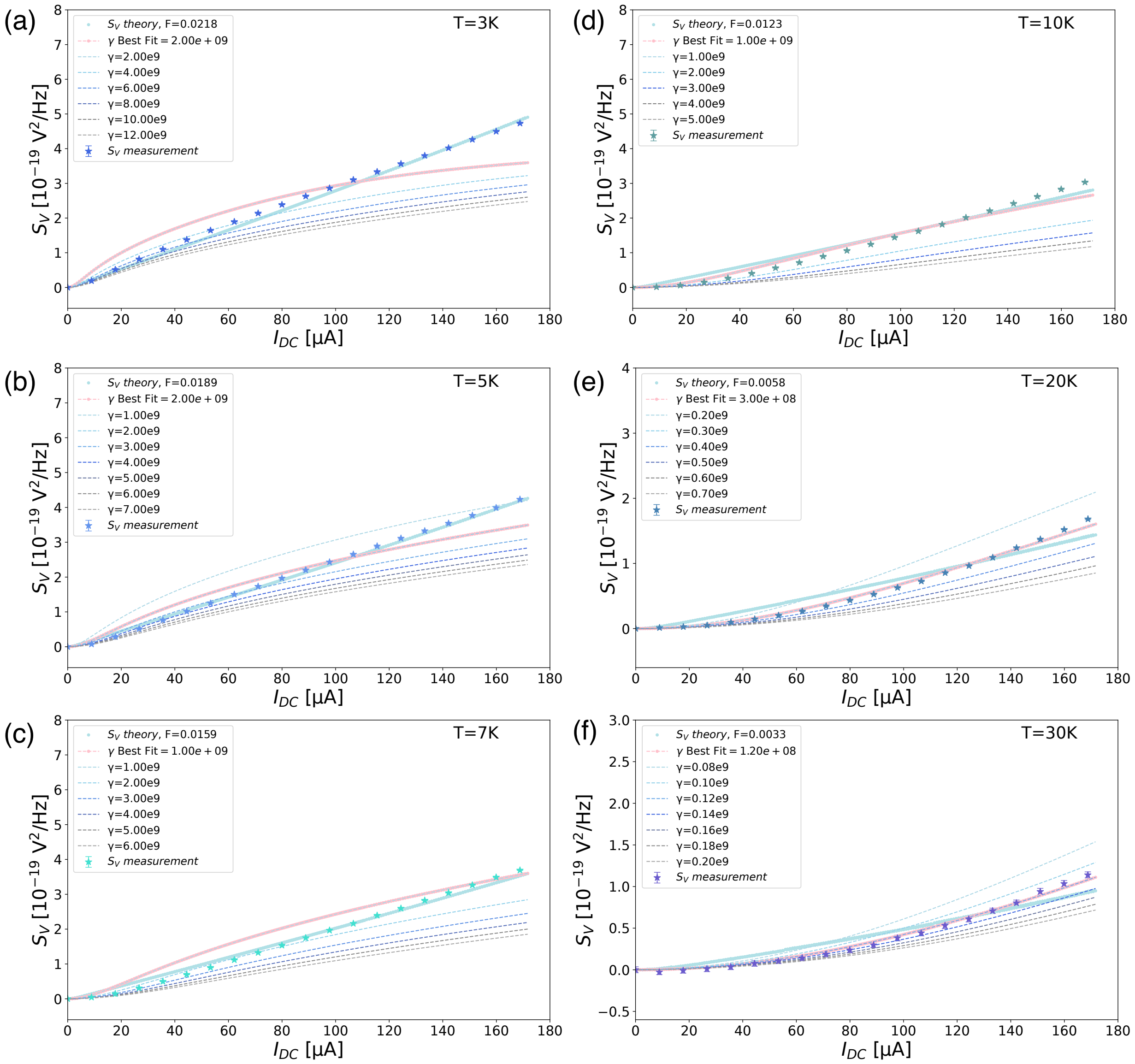

FIG 8. The excess voltage noise in a 30 μm Au nanowire on a 300 nm $SiO_2$/Si substrate. The panels show the noise (points), an illustrative fit to Eq. (7) with an effective Fano factor (solid line), and the calculated excess voltage noise as a function of bias from the $T^5$ electron thermalization with phonons model of Eq. (3), for various values of the electron-phonon energy loss parameter Γ. Panels (e, f) have a finer $y$-scale for ease of comparison. Throughout the whole temperature range, the functional form from Eq. (3) deviates systematically from the data, and the overall magnitude of the noise is considerably larger than what would be expected from just the $T^5$ model.

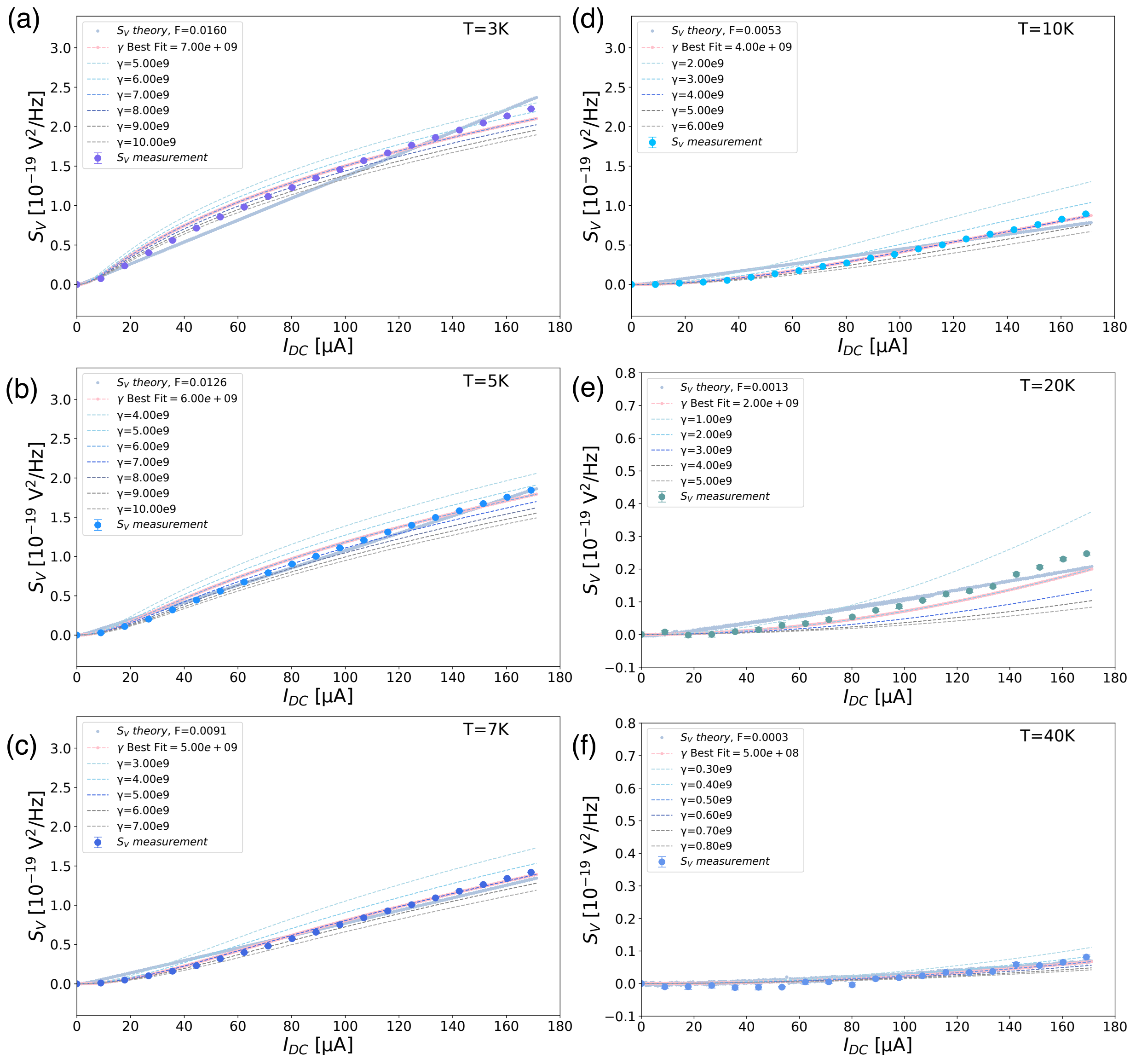

FIG 9. The excess voltage noise in a 30 μm Au nanowire on a Si substrate with only native oxide. The panels show the noise (points), an illustrative fit to Eq. (7) with an effective Fano factor (solid line), and the calculated excess voltage noise as a function of bias from the $T^5$ electron thermalization with phonons model of Eq. (3), for various values of the electron-phonon energy loss parameter Γ. Panels (e, f) have a finer *y*-scale for ease of comparison. The functional form from Eq. (3) is a better match for the data than the $SiO_2$/Si substrate case. The overall magnitude of the noise at maximum bias is smaller than that in Fig. 8 by about 2.5× at 3 K, and about 9× at 20 K. The noise is still higher than expected for the model of Eq. (3) with a fixed $\Gamma$.

Comparing Figure 9 with Fig. 8 is enlightening. When the thick oxide layer is no longer present, the functional form of $S_{V,excess}(I)$ from the electron thermalization with phonons model is a much better descriptor of the data. The 3 K data is reasonably well matched using $\Gamma \approx 7 \times 10^{9}$ $K^{-3}m^{-2}$, though this still undershoots the highest bias data, as the model again predicts a profile with greater negative curvature than the data. Up to 10 K, the best fit Γ decreases to $4 \times 10^{9}$ $K^{-3}m^{-2}$, and continues to fall slowly, ending up at $\Gamma \approx 5 \times 10^{8}$ $K^{-3}m^{-2}$ at 40 K. Again, this implies that the excess noise at 40 K is larger than what would be expected from the model of Eq. (3) with a Γ fixed at its low temperature best value. At maximum bias, the excess noise in the $SiO_2$/Si substrate device exceeds that of the Si substrate case by about 2.5× at 3 K, and 9× at 20 K. We note that modification of the e-ph power law exponent (in Eq. (3)) to 4 or 6 does not yield fits consistent with the data for a fixed Γ for either of the devices.

The sample mounting schemes and wiring for the Si and $SiO_2$/Si substrate devices are identical, and the resistances are similar ($\frac{dV}{dI}(I=0) = 525\ \Omega$ at 3 K for Si, 635 Ω for $SiO_2$). Therefore, the difference in excess noise and hence electronic heating between the two devices must result from the amorphous $SiO_2$ layer. Interestingly, the relative difference in excess noise at high bias becomes *larger* with increasing $T_0$, suggesting that this is unlikely to be from a purely interfacial boundary resistance.

The overall trends can be captured empirically by a toy model in which $\Gamma$ is assumed constant at the literature value of $\sim 5 \times 10^{9}$ $K^{-3}m^{-2}$ and the phonon temperature, $T_{ph}$, for use in Eq. (3) floats above the cryostat temperature $T_0$ in the form $T_{ph} \approx \left(T_0^2 + a(IV)\right)^{1/2}$ . Figure 10 shows examples of the quality of agreement, if $a = 1.5 \times 10^{7}$ $K^2/W$ at 5 K and $1.02 \times 10^{7}$ $K^2/W$ at 30 K. This relative change in $a$ is approximately consistent with the reported change in thermal conductivity of amorphous $SiO_2$ over this temperature range[38], further pointing at the $SiO_2$ layer as the likely root of this issue.

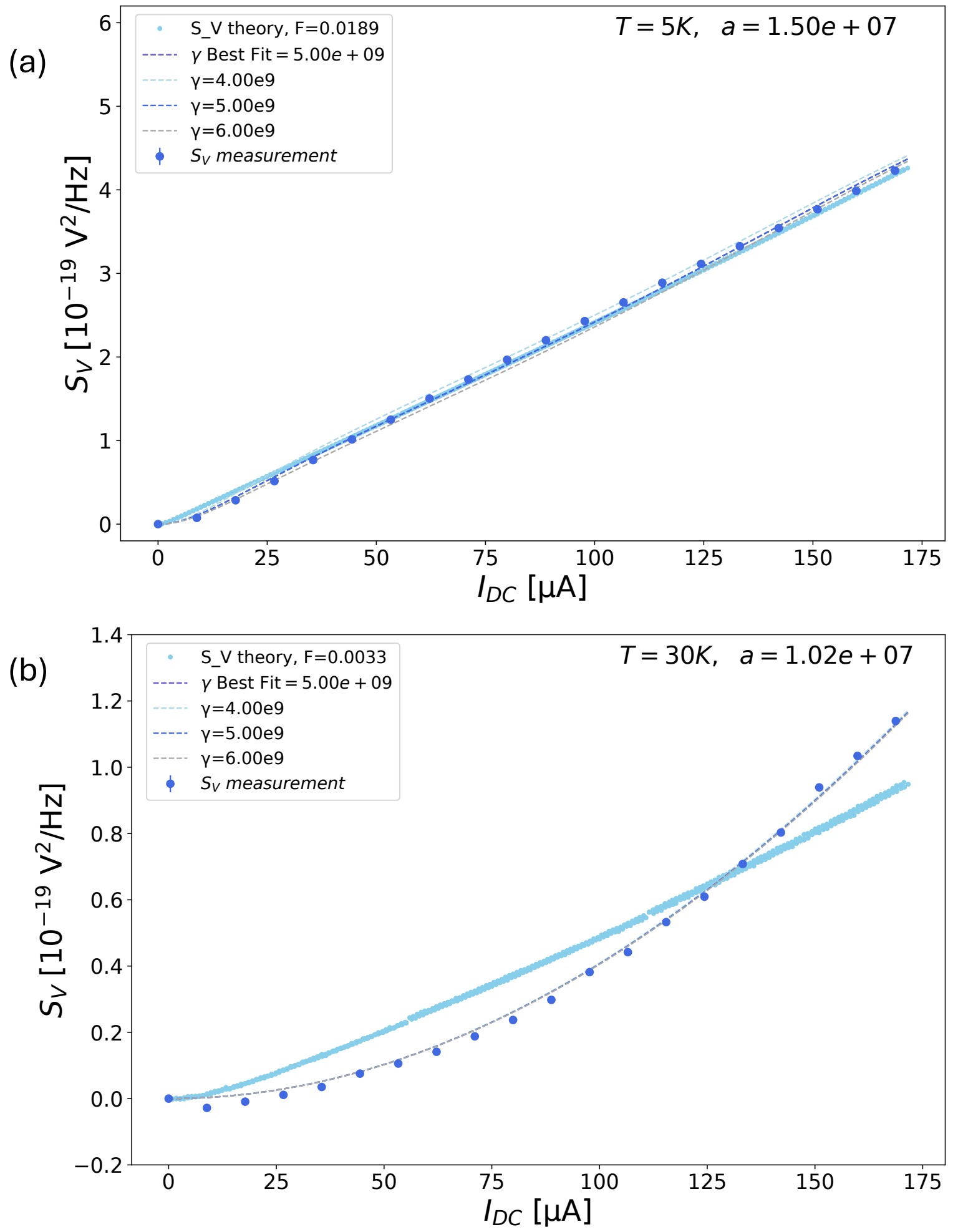

FIG 10. Comparing excess noise as a function of bias in the 30 μm wire on 300 nm $SiO_2$/Si at two temperatures (a, b) with a model that assumes the validity of the model of Eq. (3) with constant $\Gamma$, but assumes an additional thermal resistance between the phonons in the wire and the phonons in the bulk of the Si substrate, characterized by the parameter $a(T_0)$. Adding this one adjustable parameter greatly improves agreement with the data. (Solid line shows the "best fit" expected result for noise of the shot noise form in Eq. (7), purely for comparison.)

## V. CONCLUSIONS

This examination of excess noise in Au nanowires across a broad range of wire lengths and temperatures demonstrates the previously expected suppression of shot noise with increasing length and temperature[10,11], a trend that has not previously been demonstrated experimentally. In comparatively short wires ($L < 3$ μm) on 300 nm $SiO_2$/Si substrates, there is a clear suppression of shot noise with increasing length and temperature in the awkward regime when electron-phonon scattering is seemingly important but there likely is not a well-defined $T_e(x)$. The functional form of the excess noise in this regime is consistent with the shot noise form with a suppressed Fano factor fit using Eq. (7). While more involved theoretical treatments of the nonequilibrium distribution problem in such structures are possible, knowing that the data can be fit using a model with a single adjustable parameter, $F$, implies such an analysis may not constrain the underlying nonequilibrium model much. This highlights the challenge of inferring complex inelastic processes in the electrons from only the bias-dependent excess noise.

Excess noise in 30 μm long wires raises a further complication in using the bias-dependent JN noise as a diagnostic for electron-phonon energy transfer. When there is minimal interfacial oxide between the metal and the crystalline Si substrate, the bias-dependent noise at low temperatures quantitatively close to the expected functional form from Eq. (3) at the low-temperature reported value for $\Gamma$. However, at substantially elevated temperatures, the measured excess noise exceeds such expectations, even though the simplest expectation would be that phonon-based thermal boundary resistances and overall phonon transport should improve with increasing $T_0$. In contrast, when there is a 300 nm layer of amorphous $SiO_2$ between the metal and the crystalline Si substrate, the excess noise is substantially elevated relative to the Si case by an amount that increases with increasing temperature. The functional form of the noise in this case can be fit with a toy model assuming conventional e-ph scattering as in Eq. (3), but with an additional thermal resistance from the $SiO_2$ leading to an elevation of the metal phonon $T_{ph}$ above the cryostat temperature $T_0$. This agreement again serves as a reminder that with only two adjustable parameters (a temperature independent $\Gamma$ and a thermal resistance term $a$) the excess noise can be modeled reasonably well.

Another point should be considered when planning and interpreting noise measurements over broad temperature ranges and device geometries. When possible, purely crystalline structures with minimal numbers of interfaces are advantageous. A layer of $SiO_2$ is routinely used in nanostructures made from exfoliated 2D materials such as graphene and transition metal dichalcogenides because its presence improves optical contrast during stacking assembly. The presence of this layer may complicate interpretation of measurements, especially as the functional form of the excess noise can resemble shot noise expectations (Fig. 10a) even when no shot noise is involved.

**Acknowledgments**

YW, DTL, LC, MM, GJ, and DN acknowledge support for this research from DOE BES award DE-FG02-06ER46337 and acknowledge the nanofabrication infrastructure of the Rice Shared Equipment Authority. Further, DN acknowledges support from NSF award DMR-2102028 for support of the helium reliquefier measurement infrastructure. The authors thank Prof. Matthew Foster for many helpful discussions.

**Data availability**

The data for this study are available via public archive. [39]

REFERENCES

[1] J. B. Johnson, Thermal Agitation of Electricity in Conductors, Phys. Rev. **32**, 97 (1928).

[2] H. Nyquist, Thermal Agitation of Electric Charge in Conductors, Phys. Rev. **32**, 110 (1928).

[3] M. B. Weissman, 1/*f* noise and other slow, nonexponential kinetics in condensed matter, Rev. Mod. Phys. **60**, 537 (1988).

[4] W. Schottky, Regarding spontaneous current fluctuation in different electricity conductors, Ann Phys-Berlin **57**, 541 (1918).

[5] Y. M. Blanter and M. Buttiker, Shot noise in mesoscopic conductors, Phys Rep **336**, 1 (2000).

[6] K. Kobayashi and M. Hashisaka, Shot Noise in Mesoscopic Systems: From Single Particles to Quantum Liquids, J. Phys. Soc. Japan **90**, 102001 (2021).

[7] C. W. J. Beenakker and M. Büttiker, Suppression of shot noise in metallic diffusive conductors, Phys. Rev. B **46**, 1889 (1992).

[8] K. E. Nagaev, On the shot noise in dirty metal contacts, Physics Letters A **169**, 103 (1992).

[9] K. E. Nagaev, Influence of electron-electron scattering on shot noise in diffusive contacts, Phys. Rev. B **52**, 4740 (1995).

[10] A. H. Steinbach, J. M. Martinis, and M. H. Devoret, Observation of Hot-Electron Shot Noise in a Metallic Resistor, Phys Rev Lett **76**, 3806 (1996).

[11] M. Henny, S. Oberholzer, C. Strunk, and C. Schönenberger, 1/3-shot-noise suppression in diffusive nanowires, Phys. Rev. B **59**, 2871 (1999).

[12] F. C. Wellstood, C. Urbina, and J. Clarke, Hot-electron effects in metals, Phys Rev B **49**, 5942 (1994).

[13] M. Henny, H. Birk, R. Huber, C. Strunk, A. Bachtold, M. Krüger, and C. Schönenberger, Electron heating effects in diffusive metal wires, Appl. Phys. Lett. **71**, 773 (1997).

[14] M. Büttiker, Scattering theory of current and intensity noise correlations in conductors and wave guides, Phys. Rev. B **46**, 12485 (1992).

[15] H. Pothier, S. Gueron, N. O. Birge, D. Esteve, and M. H. Devoret, Energy distribution function of quasiparticles in mesoscopic wires, Phys Rev Lett **79**, 3490 (1997).

[16] A. Anthore, F. Pierre, H. Pothier, and D. Esteve, Magnetic-Field-Dependent Quasiparticle Energy Relaxation in Mesoscopic Wires, Physical Review Letters **90**, 076806 (2003).

[17] B. L. Altshuler, A. G. Aronov, and D. E. Khmelnitsky, Effects of electron-electron collisions with small energy transfers on quantum localisation, J. Phys. C: Sol. State Phys. **15**, 7367 (1982).

[18] B. Huard, A. Anthore, F. Pierre, H. Pothier, N. O. Birge, and D. Esteve, Intensity of Coulomb interaction between quasiparticles in diffusive metallic wires, Solid State Communications **131**, 599 (2004).

[19] F. Giazotto, T. T. Heikkilä, A. Luukanen, A. M. Savin, and J. P. Pekola, Opportunities for mesoscopics in thermometry and refrigeration: Physics and applications, Reviews of Modern Physics **78**, 217 (2006).

[20] B. Huard, A. Anthore, N. O. Birge, H. Pothier, and D. Esteve, Effect of Magnetic Impurities on Energy Exchange between Electrons, Physical Review Letters **95**, 036802 (2005).

[21] P. M. Echternach, M. R. Thoman, C. M. Gould, and H. M. Bozler, Electron-phonon scattering rates in disordered metallic films below 1 K, Phys Rev B **46**, 10339 (1992).

[22] J. Rammer, *Quantum Transport Theory* (Perseus Books, Reading, MA, 1998), Frontiers in Physics, p.^pp. 409-412.

[23] A. Sergeev, B. S. Karasik, M. Gershenson, and V. Mitin, Electron–phonon scattering in disordered metallic films, Physica B: Condensed Matter **316-317**, 328 (2002).

[24] A. Sergeev and V. Mitin, Electron-phonon interaction in disordered conductors: Static and vibrating scattering potentials, Phys Rev B **61**, 6041 (2000).

[25] B. Huard, H. Pothier, D. Esteve, and K. E. Nagaev, Electron heating in metallic resistors at sub-Kelvin temperature, Phys Rev B **76**, 165426 (2007).

[26] L. Chen, D. T. Lowder, E. Bakali, A. M. Andrews, W. Schrenk, M. Waas, R. Svagera, G. Eguchi, L. Prochaska, Y. Wang *et al.*, Shot noise in a strange metal, Science **382**, 907 (2023).

[27] C. Kittel and P. L. McEuen, *Kittel's Introduction to Solid State Physics, Global Edition* (Wiley, 2018).

[28] W. Ma, X. Zhang, and K. Takahashi, Electrical properties and reduced Debye temperature of polycrystalline thin gold films, Journal of Physics D: Applied Physics **43**, 465301 (2010).

[29] T. Kawamura and Y. Kato, Nonequilibrium Electron Distribution Function in a Voltage-Biased Nanowire: A Nonequilibrium Green's Function Approach, J Phys Soc Jpn **95**, 034702 (2026).

[30] E. T. Swartz and R. O. Pohl, Thermal boundary resistance, Reviews of Modern Physics **61**, 605 (1989).

[31] J. Chen, X. Xu, J. Zhou, and B. Li, Interfacial thermal resistance: Past, present, and future, Reviews of Modern Physics **94**, 025002 (2022).

[32] D. C. Glattli, P. Jacques, A. Kumar, P. Pari, and L. Saminadayar, A noise detection scheme with 10 mK noise temperature resolution for semiconductor single electron tunneling devices, Journal of Applied Physics **81**, 7350 (1997).

[33] H. E. van den Brom and J. M. van Ruitenbeek, Quantum Suppression of Shot Noise in Atom-Size Metallic Contacts, Phys Rev Lett **82**, 1526 (1999).

[34] S. Kogan, *Electronic Noise and Fluctuations in Solids* (Cambridge University Press, 2008).

[35] A. Trionfi, S. Lee, and D. Natelson, Electronic coherence in metals: Comparing weak localization and time-dependent conductance fluctuations, Phys. Rev. B **70**, 041304(R) (2004).

[36] A. Trionfi, S. Lee, and D. Natelson, Time-dependent universal conductance fluctuations in mesoscopic Au wires: Implications, Phys. Rev. B **75**, 104202 (2007).

[37] G. Cowan, *Statistical Data Analysis* (Oxford University Press, 1998).

[38] R. O. Pohl, X. Liu, and E. Thompson, Low-temperature thermal conductivity and acoustic attenuation in amorphous solids, Reviews of Modern Physics **74**, 991 (2002).

[39] The data associated with the figures in the manuscript can be accessed directly at the Zenodo data repository at https://doi.org/10.5281/zenodo.21282886.